\documentclass[showpacs,pra,twocolumn,aps]{revtex4}

\usepackage[T1]{fontenc}
\usepackage[latin1]{inputenc}
\usepackage{bm}
\usepackage{amsmath}
\usepackage{amssymb}
\usepackage{latexsym}
\usepackage{amsfonts}
\usepackage{epsfig}
\usepackage{psfrag}

\newcommand{\nn}{\nonumber\\}

\newcommand{\f}[1]{\mbox{\boldmath$#1$}}

\newcommand{\bea}{\begin{eqnarray}}
\newcommand{\ea}{\end{eqnarray}}
\newcommand{\eea}{\end{eqnarray}}
\newcommand{\ord}{\,{\cal O}}

\begin{document}

\title{Effect of fluctuations on the superfluid-supersolid phase
  transition on the lattice} 

\author{Ralf Sch\"utzhold$^{1,2}$, Michael Uhlmann$^{1,3}$, and Uwe
  R.~Fischer$^4$}  

\affiliation{$^1$Institut f\"ur Theoretische Physik, Technische
  Universit\"at Dresden, D-01062 Dresden, Germany\\
$^2$Fachbereich Physik, Universit\"at Duisburg-Essen, D-47048
  Duisburg, Germany\\ 
$^3$Department of Physics, Australian National University, Canberra,
  ACT 0200, Australia\\
$^4$Eberhard-Karls-Universit\"at T\"ubingen,
Institut f\"ur Theoretische Physik\\
Auf der Morgenstelle 14, D-72076 T\"ubingen, Germany} 

\begin{abstract}
We derive a controlled expansion into mean field plus fluctuations 
for the extended Bose-Hubbard model, involving interactions with many
neighbors on an arbitrary periodic lattice, and study the
superfluid-supersolid phase transition. 
Near the critical point, the impact of (thermal and quantum)
fluctuations on top of the mean field 
grows, which entails striking effects, such as negative superfluid
densities and thermodynamical instability of the superfluid phase --
earlier as expected from mean-field dynamics.  
% ###
We also predict the existence of long-lived ``supercooled'' states 
with anomalously large quantum fluctuations.
\end{abstract}


\maketitle

%%%%%%%%%%%%%%%%%%%%%%%%%%%%%%%%%%%%%%%%%%%%%%%%%%%%%%%%%%%%%%%%%%%%%%%%%%%%%%
%%%%%%%%%%%%%%%%%%%%%%%%%%%%%%%%%%%%%%%%%%%%%%%%%%%%%%%%%%%%%%%%%%%%%%%%%%%%%%
%%%%%%%%%%%%%%%%%%%%%%%%%%%%%%%%%%%%%%%%%%%%%%%%%%%%%%%%%%%%%%%%%%%%%%%%%%%%%%
%%%%%%%%%%%%%%%%%%%%%%%%%%%%%%%%%%%%%%%%%%%%%%%%%%%%%%%%%%%%%%%%%%%%%%%%%%%%%%
%%%%%%%%%%%%%%%%%%%%%%%%%%%%%%%%%%%%%%%%%%%%%%%%%%%%%%%%%%%%%%%%%%%%%%%%%%%%%%
%%%%%%%%%%%%%%%%%%%%%%%%%%%%%%%%%%%%%%%%%%%%%%%%%%%%%%%%%%%%%%%%%%%%%%%%%%%%%%

%%%%%%%%%%%%%%%%%%%%%%%%%%%%%%%%%%%%%%%%%%%%%%%%%%%%%%%%%%%%%%%%%%%%%%%%%%%%%%
%%%%%%%%%%%%%%%%%%%%%%%%%%%%%%%%%%%%%%%%%%%%%%%%%%%%%%%%%%%%%%%%%%%%%%%%%%%%%%
\section{Introduction} 
%%%%%%%%%%%%%%%%%%%%%%%%%%%%%%%%%%%%%%%%%%%%%%%%%%%%%%%%%%%%%%%%%%%%%%%%%%%%%%
%%%%%%%%%%%%%%%%%%%%%%%%%%%%%%%%%%%%%%%%%%%%%%%%%%%%%%%%%%%%%%%%%%%%%%%%%%%%%%

The question of whether macroscopic quantum coherence can prevail in
the presence of periodic order, ultimately leading to the existence of
a supersolid, has been intensely debated since five
decades \cite{Penrose,Andreev,LeggettSuperSolid}.  
Of late, this topic has seen a renewed surge of interest, partly due
to the observations in \cite{SSNature} indicating potential signatures
of a supersolid phase of $^4$He. 
Because of the inherent complexity of $^4$He, it is useful to gain
further understanding by studying supersolid phases in other 
systems -- such as the Bose-Hubbard model, which can be realized
experimentally via cold bosonic atoms in optical lattices
\cite{Jaksch}. 
For on-site interactions only, the phase diagram at $T=0$ contains 
the superfluid and the Mott insulator state \cite{Xu,Greiner}. 
Adding interactions across sites (next nearest or higher) in a 
so-called extended Bose-Hubbard model, a superfluid-supersolid phase 
transition may occur \cite{Batrouni,Scarola,Kovrizhin,Goral,Wessel}
in addition to further Mott-like phases. 
Here the term supersolid is associated to an order parameter 
(with a well-defined phase) which is, in contrast to the homogeneous
superfluid ground state, not the same for all lattice sites, but
periodically modulated. 
At the heart of supersolid formation is the generic phenomenon that an 
instability towards density modulations occurs if the excitation
spectrum dips below zero for a finite wavevector. 

Within mean-field theory, i.e., neglecting all fluctuations, 
properties of the supersolid phase were studied in \cite{Josserand}. 
However, the evanescent excitation energies at the transition
suggest that (thermal and quantum) fluctuations should play an
important role near the critical point. 
The impact of these fluctuations can be taken into account with
quantum Monte Carlo simulations, see, e.g., \cite{Batrouni,Wessel}.  
Despite the strength of this method, these simulations
are always restricted to a specific (low-dimensional) lattice of
finite size and a small sample of the full Hilbert space. 
In the following, we consider an arbitrary periodic lattice and
develop an analytic expansion into mean field plus fluctuations 
where the size of the fluctuations and the validity of the 
expansion is {\em controlled} by a small parameter. 
Therefore, our derivation is complementary to other numerical and
analytical approaches using for example %the 
duality to vortex field theory %employed in 
\cite{Burkov+Balents}. 
To the end of devising a controlled
mean field expansion, we begin by introducing
the concept of weighted operator sums in the next section.  

%%%%%%%%%%%%%%%%%%%%%%%%%%%%%%%%%%%%%%%%%%%%%%%%%%%%%%%%%%%%%%%%%%%%%%%%%%%%%%
%%%%%%%%%%%%%%%%%%%%%%%%%%%%%%%%%%%%%%%%%%%%%%%%%%%%%%%%%%%%%%%%%%%%%%%%%%%%%%
\section{Weighted operator sums} %###
%%%%%%%%%%%%%%%%%%%%%%%%%%%%%%%%%%%%%%%%%%%%%%%%%%%%%%%%%%%%%%%%%%%%%%%%%%%%%%
%%%%%%%%%%%%%%%%%%%%%%%%%%%%%%%%%%%%%%%%%%%%%%%%%%%%%%%%%%%%%%%%%%%%%%%%%%%%%%

We consider the extended Bose-Hubbard model on an arbitrary lattice as
described by the Hamiltonian 
\bea
\label{Hamiltonian}
\hat H
=
\sum_{\alpha\beta}
\left(
T_{\alpha\beta}\,
\hat a_\alpha^\dagger\hat a_\beta
+
\frac12V_{\alpha\beta}\,
\hat a_\alpha^\dagger
\hat a_\beta^\dagger
\hat a_\alpha
\hat a_\beta
\right)
\,,
\ea
where $\alpha,\beta$ label the lattice sites and 
$\hat a_\alpha^\dagger,\hat a_\beta$ are the associated bosonic 
creation/annihilation operators. 
The kinetic term is determined by the hopping matrix $T_{\alpha\beta}$ 
and the interaction part by $V_{\alpha\beta}$. 
(Both matrices are real and symmetric.)
Since the above Hamiltonian cannot be diagonalized analytically, we
have to employ some approximations. 
To this end, we introduce the concept of weighted operator sums
defined via 
\bea
\label{weighted}
\hat X_{S}[f]=\frac{1}{|S|}
\sum_{\alpha\in S}
f_\alpha(\hat a_\alpha^\dagger,\hat a_\alpha)
\,,
\ea
with a set $S\subset\mathbb N$ of $|S|$ elements $\alpha\in S$ and a
function $f_\alpha$ of order one, i.e., which does not scale with
$|S|$. 
Hence the limit $\lim_{|S|\to\infty}\hat X_{S}[f]$ exists 
(in an appropriate sense, e.g., as a weak limit) while all single
addends are suppressed by $1/|S|$ for large $|S|$. 
Examples for the form (\ref{weighted}) include all (local) one-site 
operators such as   
$\hat X_{\{\alpha\}}[\f{1}]=\hat a_\alpha$ for $|S|=1$  
as well as the (global) Fourier components 
$\sum_\alpha\hat a_\alpha\exp\{ik\alpha\}/L=\hat a_{k}/\sqrt{L}$
with $|S|=L$ being the total number of sites for a one-dimensional
chain $S=[1,L]$.  
Now, considering the commutator between two such weighted operator sums 
\bea
\label{commutator} 
\left[\hat X_{S}[f],\hat X_{S'}[f']\right]
=
\frac{|S \cap S'|}{|S|\times|S'|}
\,
\hat X_{S \cap S'}[f'']
\,,
\ea
with $f_\alpha''(\hat a_\alpha^\dagger,\hat a_\alpha)=
[f_\alpha(\hat a_\alpha^\dagger,\hat a_\alpha),
f_\alpha'(\hat a_\alpha^\dagger,\hat a_\alpha)]$, 
we find that they are suppressed for large $|S|$ due to 
$|S \cap S'|\leq{\rm min}\{|S|,|S'|\}$.
Hence the limit $\lim_{|S|\to\infty}\hat X_{S}[f]$ commutes with all
other weighted operator sums (including all local operators) and can
thus be approximated by a c-number within the relevant Hilbert space
generated by weighted operator sums acting on the ground 
(or thermal) state. 
This motivates the following asymptotic expansion for large $|S|\gg1$ 
\bea
\label{asymptotic}
\hat X_{S}[f]
=
\hat C_0[f]+\frac{\hat C_{1/2}[f]}{\sqrt{|S|}}+
\frac{\hat C_{1}[f]}{|S|}+\dots 
\,,
\ea
where the leading term $\hat C_0[f]$ can be approximated by a c-number 
and the sub-leading operators $\hat C_{1/2}[f]$ and $\hat C_{1/2}[f']$ 
generate the commutator (\ref{commutator}), of order $1/|S|$. 

Applying the concept of weighted operator sums to operators like 
$\hat X_{\Sigma}[\f{1}]=\sum_\beta\hat a_\beta/L$ or other Fourier
components, we arrive at the mean-field expansion 
\bea
\label{mean-field}
\hat a_\alpha
=
\psi_\alpha+\hat\chi_\alpha+\ord(1/\sqrt{L})
\,,
\ea
where $\psi_\alpha$ denotes the mean field and corresponds to the
leading parts $\hat C_0[f]$ in Eq.~(\ref{asymptotic}) while the
fluctuations $\hat\chi_\alpha$ with $\langle\hat\chi_\alpha\rangle=0$
incorporate the non-commuting remainders.
Note that (in contrast to \cite{unser}) the filling 
$n_\alpha=\langle\hat a_\alpha^\dagger\hat a_\alpha\rangle=
|\psi_\alpha^2|+\langle\hat\chi_\alpha^\dagger\hat\chi_\alpha\rangle$ 
is here not assumed to be large; $|\psi_\alpha^2|$ is the condensate
part and $\langle\hat\chi_\alpha^\dagger\hat\chi_\alpha\rangle$ is the 
remaining thermal or quantum depletion.
Hence the fluctuations $\hat\chi_\alpha$ are not necessarily small
compared to the mean field $\psi_\alpha$: 
e.g., for half-filling $n_\alpha=1/2$, the variance is obviously of
order one.  
In order to simplify the full equation of motion derived from
(\ref{Hamiltonian}) [$\hbar=1$]
\bea
\label{full}
i\partial_t\hat a_\alpha
=
\sum_{\beta}
\left(
T_{\alpha\beta}\hat a_\beta
+
V_{\alpha\beta}
\hat n_\beta\hat a_\alpha
\right)
\,,
\ea
we assume that the interaction $V_{\alpha\beta}$ involves a large
number $D\gg1$ of sites $\beta$ on a roughly equal footing. 
This could be the case, for example, for 
long-range interactions or for a large
number of spatial dimensions. 
For normalized potentials
$\sum_{\beta}V_{\alpha\beta}\equiv V_\Sigma=\ord(1)$, 
we may then apply the concept of weighted operator sums
(\ref{weighted}) to the term $\sum_{\beta}V_{\alpha\beta}\hat n_\beta$
and obtain 
\bea
\sum_{\beta}V_{\alpha\beta}\hat n_\beta=
\sum_{\beta}V_{\alpha\beta}\langle\hat
n_\beta\rangle+\ord(1/\sqrt{D})
\ea
%
%###
%
%$\sum_{\beta}V_{\alpha\beta}\hat n_\beta=
%\sum_{\beta}V_{\alpha\beta}\langle\hat
%n_\beta\rangle+\ord(1/\sqrt{D})$
from Eq.~(\ref{asymptotic}). 
However, one must be careful: simply replacing $\hat n_\beta$ by
$n_\beta$ in (\ref{full}), we would lose the phonon modes. 
The sub-leading term $\ord(1/\sqrt{D})$ can only be neglected if there
is no {\em other} small (or large) term involved. 
This is precisely the case for modes with long wavelengths %reaching
over many lattice sites, where the sum over 
$T_{\alpha\beta}\hat a_\beta$, for example, is also very small and
hence the $\ord(1/\sqrt{D})$ contributions become relevant. 
In order to describe long-wavelength modes correctly, we insert
Eq.~(\ref{mean-field}) into Eq.~(\ref{full}) to
obtain the Gross-Pitaevski\v\i\/ equation 
\bea
\label{GP}
i\partial_t\psi_\alpha
=
\sum_{\beta}
\left(
T_{\alpha\beta}\psi_\beta
+
V_{\alpha\beta}
\left[
|\psi_\beta|^2
+
\langle\hat\chi_\beta^\dagger\hat\chi_\beta\rangle
\right]
\psi_\alpha
\right)
\,,
\ea
where we have replaced 
$\sum_{\beta}V_{\alpha\beta}\hat\chi_\beta^\dagger\hat\chi_\beta$ 
by its expectation value according to the above arguments, 
plus the remaining fluctuation part
%
%###
%
\bea
\label{pre-BdG}
i\partial_t\hat\chi_\alpha
&=&
\sum_{\beta}
\left(
T_{\alpha\beta}\hat\chi_\beta
+
V_{\alpha\beta}
\left[|\psi_\beta|^2+
\langle\hat\chi_\beta^\dagger\hat\chi_\beta\rangle\right]
%\hat\chi_\beta^\dagger\hat\chi_\beta\right]
\hat\chi_\alpha
\right.
\nn
&&
+
V_{\alpha\beta}
\left.
\left[
\psi_\beta^*\hat\chi_\beta
+\psi_\beta\hat\chi_\beta^\dagger
%+\hat\chi_\beta\hat\chi_\beta^\dagger
\right]
(\psi_\alpha+\hat\chi_\alpha)
\right)
\,.
\ea
Again, the second line is suppressed by $\ord(1/\sqrt{D})$ and will
only be relevant for long-wavelength modes, which involve a sum over
many sites $\alpha$.
In this case, however, the c-number term $\sum_\alpha\psi_\alpha$ will
dominate the fluctuation term $\sum_\alpha\hat\chi_\alpha$ in view of 
Eq.~(\ref{asymptotic}) and hence we may approximate the bracket
$(\psi_\alpha+\hat\chi_\alpha)$ in the second line by $\psi_\alpha$,
arriving at a linear operator equation
\begin{multline}
\label{BdG}
i\partial_t\hat\chi_\alpha
=
\sum_{\beta}
\left(
T_{\alpha\beta}\hat\chi_\beta
+
V_{\alpha\beta}
\left[|\psi_\beta|^2+
\langle\hat\chi_\beta^\dagger\hat\chi_\beta\rangle\right]
\hat\chi_\alpha
\right.
\\
%\nn&&
+
V_{\alpha\beta}
\left.
\left[
\psi_\beta^*\hat\chi_\beta
+\psi_\beta\hat\chi_\beta^\dagger
\right]\psi_\alpha
\right)
+\ord (1/{\sqrt{D}} )
\,,
\end{multline} 
%\ea
%
which corresponds to the Bogoliubov-de~Gennes equations for the
fluctuations.  
Note that the approximation from Eq.~(\ref{pre-BdG}) to
Eq.~(\ref{BdG}) neglects the exchange of particles between the
condensate $|\psi_\alpha^2|$ and the thermal or quantum depletion 
$\langle\hat\chi_\alpha^\dagger\hat\chi_\alpha\rangle$.
This exchange is governed by the sub-leading term 
$\sum_{\beta}V_{\alpha\beta}\psi_\beta^*
\langle\hat\chi_\beta\hat\chi_\alpha\rangle$, which could be added to  
Eq.~(\ref{GP}). 

%%%%%%%%%%%%%%%%%%%%%%%%%%%%%%%%%%%%%%%%%%%%%%%%%%%%%%%%%%%%%%%%%%%%%%%%%%%%%%
%%%%%%%%%%%%%%%%%%%%%%%%%%%%%%%%%%%%%%%%%%%%%%%%%%%%%%%%%%%%%%%%%%%%%%%%%%%%%%
\section{Quasiparticle modes} %###
%%%%%%%%%%%%%%%%%%%%%%%%%%%%%%%%%%%%%%%%%%%%%%%%%%%%%%%%%%%%%%%%%%%%%%%%%%%%%%
%%%%%%%%%%%%%%%%%%%%%%%%%%%%%%%%%%%%%%%%%%%%%%%%%%%%%%%%%%%%%%%%%%%%%%%%%%%%%%

In order to introduce quasiparticle modes, we assume translational
invariance, i.e., that $T_{\alpha\beta}$ and $V_{\alpha\beta}$ only
depend on the distance $\alpha-\beta$ and that the condensate density
is homogeneous $|\psi_\alpha|=|\psi|$.
Nevertheless, we may still have a constant phase gradient $\eta$ in
our sample, i.e., we set 
$\psi_\alpha=|\psi|\exp\{-i\mu t+i\eta\alpha\}$.
In this case, we may diagonalize Eq.~(\ref{BdG}) via a Fourier
transformation
\bea
i\partial_t\hat\chi_k
=
\left(
T_{k+\eta}+V_\Sigma n+V_k|\psi^2|
\right)
\hat\chi_k 
+
V_k\psi^2\hat\chi_{-k}^\dagger
\,.
\ea
%
%###
%$i\partial_t\hat\chi_k=
%\left(
%T_{k+\eta}+V_\Sigma n+V_k|\psi^2|
%\right)
%\hat\chi_k 
%+
%V_k\psi^2\hat\chi_{-k}^\dagger$. 
%
Note that in more than one spatial dimension, $\alpha$ and $\beta$
as well as $k$ and $\eta$ will be multi-indices 
(labeling the real and the inverse lattice, respectively)  
and $\eta\alpha$ is a scalar product. 
Assuming reflection invariance $T_k=T_k^*=T_{-k}$ and
$V_k=V_k^*=V_{-k}$ for the lattice, we see that this symmetry $k\to-k$
is broken for the  modes $\hat\chi_k$ by the phase gradient
$\eta$. 
The quasiparticle Hamiltonian 
\bea 
\label{phonon}
\hat H_\chi 
&=&  
\sum_{k} 
\left\{
\hat\chi^\dagger_k 
\left(T_{k+\eta} +V_\Sigma n+|\psi|^2V_k \right)
\hat\chi_k 
\phantom{\frac12}
\right.
\nn
&&
\left.
+\frac{V_k}{2}
\left(\psi^2\hat\chi_k^\dagger \hat\chi_{-k}^\dagger + {\rm h.c.}\right)
\right\}
=  
\sum_{k} \omega_k^+ \hat b_k^\dagger\hat b_k 
\,,
\ea 
can be diagonalized via the Bogoliubov transformation 
$\hat\chi_k=u_k\hat b_k+v_{k}\hat b^\dagger_{-k}$
with $|u_k^2|-|v_k^2|=1$.
This yields the   Bogoliubov coefficients 
%###
\bea 
u_k & = & \frac{1}{1-l_k^2}\,, \quad v_k=\frac{l_k}{1-l_k^2}\,, 
\nn 
l_k & = & \sqrt{w_k^2+2w_k}-1-w_k\,, \quad w_k=\frac{\bar T_k}{V_k|\psi^2|}
\,. 
%\nn
\label{coeff}
\ea
For $w_k=-2$, the coefficients diverge due to $l_k=1$ 
(leading to the instability for $\eta=0$, to be discussed below). 
The quasiparticle frequencies obey the dispersion relation 
($T_{k=0}=0$)
\bea
\label{dispersion}
\omega_k^\pm
=
\frac12(T_{k+\eta}-T_{k-\eta}) 
\pm
\sqrt{\bar T_k^2+2|\psi|^2 V_k \bar T_k}
\,,
\ea
where $\bar T_k = (T_{k+\eta}+T_{k-\eta})/2$ 
and thus the branches are connected by $\omega_k^+=-\omega_{-k}^-$. 

In the continuum limit, i.e., for small $k\ll1$, we may approximate
$T_k\approx k^2/(2m)$ due to $T_k=T_k^*=T_{-k}$ and $T_{k=0}=0$ 
with the mass $m$ being determined by the hopping rates.  
For small phase gradients $\eta\ll k\ll1$, we then reproduce the usual
Galilei shift 
%###
\bea
(\omega^\pm_k+vk)^2=|\psi^2|V_k\,\frac{k^2}{m}+\frac{k^4}{(2m)^2}
\,,
\ea
%
%$(\omega^\pm_k+vk)^2=|\psi^2|V_kk^2/m+k^4/(2m)^2$, 
where $v=\eta/m$ is superfluid velocity.
Now, even for purely positive $V_{\alpha\beta}$, the Fourier transform  
$V_k$ may become negative for some $k$ and hence the dispersion
relation may develop dips 
(similar to the roton dip in superfluid $^4\!$He). 
If $V_k$ is sufficiently negative (compared to $T_k$), the dispersion
curve $\omega_k$ may even dive below zero.
Ignoring the fluctuations discussed below, the onset of instability, 
$\omega_k=0$, marks the end of the (homogeneous) superfluid phase and
the beginning of the supersolid phase where $|\psi_\alpha|$ is
periodic, i.e., inhomogeneous. 
The phase gradient $\eta$ favors the supersolid phase, i.e., the
transition superfluid $\to$ supersolid occurs earlier for
non-vanishing $\eta$.
% \cite{growth}. 
For $\eta=T=0$, the frequencies $\omega_{k=\pm k_*}$ at the roton
wavenumber become imaginary beyond the critical point and hence 
these modes start to grow exponentially.  
For $\eta> 0$ and $T=0$, the transition occurs earlier and and is
slower since the frequency $\omega_{k=+k_*}$ becomes negative, but not
imaginary. 
Hence only the coupling to some environment (fixed by the lattice) 
induces an instability of these quasiparticle modes.
On the other hand, the depletion 
$\langle\hat\chi_\alpha^\dagger\hat\chi_\alpha\rangle$ 
due to thermal or quantum fluctuations favors the superfluid phase
since it reduces (for a fixed filling $n$) the condensate fraction 
$|\psi^2|$ and thus weakens the term $|\psi^2|V_k$ in
Eq.~(\ref{dispersion}) responsible for the roton dip. 
Ergo, heating up the supersolid state may yield the superfluid phase
(as long as the condensate does not disappear altogether $\psi=0$),
which will become important for the discussion of 
``supercooled'' states we turn to in section \ref{supercooled}.

%%%%%%%%%%%%%%%%%%%%%%%%%%%%%%%%%%%%%%%%%%%%%%%%%%%%%%%%%%%%%%%%%%%%%%%%%%%%%%
%%%%%%%%%%%%%%%%%%%%%%%%%%%%%%%%%%%%%%%%%%%%%%%%%%%%%%%%%%%%%%%%%%%%%%%%%%%%%%
\section{Superfluid density} 
%%%%%%%%%%%%%%%%%%%%%%%%%%%%%%%%%%%%%%%%%%%%%%%%%%%%%%%%%%%%%%%%%%%%%%%%%%%%%%
%%%%%%%%%%%%%%%%%%%%%%%%%%%%%%%%%%%%%%%%%%%%%%%%%%%%%%%%%%%%%%%%%%%%%%%%%%%%%%

Now let us study the response of the system to a small phase gradient
$\hat a_\alpha\to\hat a_\alpha\exp\{i\eta\alpha\}$, which determines
the superfluid fraction. 
The interaction part
$\frac12 V_{\alpha\beta}\,\hat a_\alpha^\dagger\hat a_\beta^\dagger
\hat a_\alpha\hat a_\beta$ of the Hamiltonian (\ref{Hamiltonian}) 
does not change, but the kinetic term yields 
%###
\bea
\frac{\partial\hat H}{\partial\eta}
=
-i\sum_{\alpha\beta}T_{\alpha\beta}
(\alpha-\beta)\hat a_\alpha^\dagger\hat a_\beta
\,,
\ea
%
%$\partial\hat H/\partial\eta=-i\sum_{\alpha\beta}T_{\alpha\beta}
%(\alpha-\beta)\hat a_\alpha^\dagger\hat a_\beta$, 
which is a measure for the total current 
$\partial\hat H/\partial\eta\propto\hat J$, cf.~the Fourier expansion
in (\ref{phonon}). 
E.g., for $T_{\alpha\beta}\propto
\delta_{\alpha,\beta+1}+\delta_{\alpha,\beta-1}-2\delta_{\alpha,\beta}$ 
\cite{unser}, we get the usual expression 
$\hat J\propto i(\hat a_{\alpha+1}^\dagger\hat a_\alpha-{\rm h.c.})$. 
In the continuum limit of $T_k\approx k^2/(2m)$, we obtain 
\bea
\hat J\propto 
\frac{\partial\hat H}{\partial\eta}=
\frac{1}{m}
\sum_k
\left(\eta|\psi^2|+k\,\hat\chi_k^\dagger\hat\chi_k\right)
\to
J_\psi+\hat J_\chi 
\,,
\ea
where the first term $\eta|\psi^2|$ in the bracket is the condensate
(i.e., mean-field) contribution $J_\psi$ and the second one,  
$\hat J_\chi$, stems from the fluctuations. 
Inserting the Bogoliubov transformation 
$\hat\chi_k=u_k\hat b_k+v_{k}\hat b^\dagger_{-k}$, we find 
\bea
\label{depletion}
\langle\hat\chi_k^\dagger\hat\chi_k\rangle_0
=
%\langle\hat\chi_\alpha^\dagger\hat\chi_\alpha\rangle_0=
%\frac{1}{2}%\sum_k
%\left(
\frac{\bar T_k+|\psi^2|V_k}{2\sqrt{\bar T_k^2+2\bar T_k|\psi^2|V_k}}
-\frac{1}{2}
%\right)
=
\langle\hat\chi_{-k}^\dagger\hat\chi_{-k}\rangle_0
\,,
\ea
i.e., the expectation value of the fluctuation part in the ground
state vanishes $\langle\hat J_\chi\rangle_0=0$. 
Even though the quasiparticle frequencies are different in opposite
directions for a non-vanishing phase gradient $\eta$, 
$\omega_k\neq\omega_{-k}$, 
the Bogoliubov coefficients are still symmetric 
$|u_k|=|u_{-k}|$ and $|v_k|=|v_{-k}|$. 
Because of the symmetry $l_k=l_{-k}$, from Eq.\,\eqref{coeff}, quantum
depletion does not contribute to the current \cite{non-adiabatic}. %### 
%
%Note, however, that an opposite current could also be generated by
%pure quantum effects due to non-equilibrium phenomena: approaching the
%transition too fast, we get non-adiabatic excitations 
%$\langle\hat b_k^\dagger\hat b_k\rangle>0$ which are stronger for
%small $\omega_k$. 

In a thermal ensemble, as described by the density matrix 
$\hat\varrho=\exp\{-\hat H_\chi/T\}/Z$, however, quasiparticle modes
with $\omega_k\neq\omega_{-k}$ will have
different occupation numbers and hence we do get a contribution to the
total flux from the fluctuations 
$\langle\hat J\rangle\propto\sum_k
(\eta|\psi^2|+k\,\langle\hat b_k^\dagger\hat b_k\rangle)$. 
Clearly, near the superfluid-supersolid phase transition, the major 
contributions occur around the roton minima at $\pm k_*$. 
Here, we consider for simplicity 
one spatial dimension only, but the main results
apply to higher dimensions as well. 
Let us first study the case $\eta=0$. 
In the continuum limit $k\ll1$, we may use a Taylor expansion 
%###
\bea
\omega_k^2=2T_k|\psi^2|V_k+T_k^2\approx\omega_*^2+\gamma^2(k-k_*)^2
\ea
%
%$\omega_k^2=2T_k|\psi^2|V_k+T_k^2\approx\omega_*^2+\gamma^2(k-k_*)^2$ 
around the roton minimum at the critical wavenumber $k_*\ll1$, where
$\gamma$ is the curvature of the roton dip. % 
Approaching the phase transition corresponds to the limit
$\omega_*^2\to0$ and for small $\omega_*$ with $\omega_*\ll T$,  
the leading term scales as 
\bea
\label{thermal}
\frac{1}{L}\sum_{k}
\langle\hat b_k^\dagger\hat b_k\rangle=
\ord\left(\frac{T\ln\omega_*}{{\gamma}}\right)
\,.
\ea
At the critical point $\omega_*=0$, the $k$-integral over the thermal
distribution 
$\langle\hat b_k^\dagger\hat b_k\rangle\approx T/\omega_k$ becomes
weakly divergent near the roton dip at $k_*$ where 
$\omega_k\approx\gamma|k-k_*|$, leading to the logarithmic
singularity $\ln\omega_*$. 

Now, adding a small phase gradient, one roton minimum is lifted and 
the other one approaches the $\omega=0$ axis even closer. 
Hence the thermal quasiparticle occupation numbers 
$\langle\hat b_k^\dagger\hat b_k\rangle$ react in opposite ways and
induce a net current -- which is opposite to the condensate flux 
$\eta|\psi^2|$. 
The change due to $\omega_*\to\omega_*\pm vk_*$ scales as 
\bea
\langle\hat J_\chi\rangle
\propto
\frac{1}{L}\sum_{k}
k\,\langle\hat b_k^\dagger\hat b_k\rangle
=
\ord\left(\frac{Tvk_*^2}{\omega_*{\gamma}}\right)
\,.
\ea
For small enough $\omega_*$ or, alternatively, for large enough
temperatures $T > T_{\rm cr}=\ord(\omega_*m\gamma|\psi^2|/k_*^2)$, 
the current induced by the thermal fluctuations can easily compensate
the condensate (mean-field) contribution $\eta|\psi^2|$.
Thus, the superfluid fraction can be significantly reduced -- and may
even become negative (which is also occurring in $\pi$-Josephson 
junctions \cite{Bulaevskii}), i.e., the phase gradient $\eta$ entails
a net current $\langle\hat J\rangle$ in the {\em opposite} direction. 

Such a negative superfluid density induces a thermodynamical 
instability \cite{Sengupta}:
As discussed before, the current $\langle\hat J\rangle$ is a measure
for the response of the system to a phase gradient 
$\langle\partial\hat H/\partial\eta\rangle$. 
Inserting the canonical ensemble $\hat\varrho=\exp\{-\hat H/T\}/Z$, we
see that the expectation value 
$\langle\partial\hat H/\partial\eta\rangle=\partial F/\partial\eta$ 
equals the change of the free energy 
$F=E-TS=\langle\hat H\rangle+T\langle\ln\hat\varrho\rangle$.
Since a stable equilibrium state in an isothermal environment
corresponds to a minimum of the free energy, a negative superfluid
density $\partial \langle\hat J\rangle /\partial \eta <0$ 
shows that the system is unstable
against the spontaneous generation of local phase gradients
(since $\eta=0$ is a maximum of the free energy).

Note that a negative superfluid density does not require a large
thermal depletion: as we may infer from Eq.~(\ref{thermal}), the
thermal occupation number 
$\sum_{k}\langle\hat b_k^\dagger\hat b_k\rangle$
scales merely logarithmically with $\omega_*$ and hence it can be much 
smaller than the condensate fraction $|\psi^2|$ 
(e.g., for $T\ll\gamma$ and $k_*\ll1\leadsto k_*\ln\omega_*<1$). 
Of course, in addition to thermal occupation, the condensate is also
depleted by quantum effects.
This quantum depletion survives at zero temperatures and is given by 
Eq.~\eqref{depletion} via 
$\langle\hat\chi_\alpha^\dagger\hat\chi_\alpha\rangle=
\sum_k\langle\hat\chi_k^\dagger\hat\chi_k\rangle/L$.
%
%(at $\eta=0$ and $T=0$) 
%
% 
%\bea
%\langle\hat\chi_\alpha^\dagger\hat\chi_\alpha\rangle_0=
%\frac{1}{2L}\sum_k
%\left(
%\frac{T_k+|\psi^2|V_k}{\sqrt{T_k^2+2T_k|\psi^2|V_k}}-1
%\right)
%\,.
%\ea
%
%For non-zero $\eta$, the only change consists in replacing $T_k$ by 
%$\bar T_k = (T_{k+\eta}+T_{k-\eta})/2$, which shows 
%Note that the quantum depletion is an even function of $\eta$, i.e.,
%it does not contribute to the superfluid density 
%
With the same approximations as in Eq.~(\ref{thermal}), we get again
merely a logarithmic dependence 
\bea
\langle\hat\chi_\alpha^\dagger\hat\chi_\alpha\rangle_0=
\ord\left(\frac{k_*^2\ln\omega_*}{m\gamma}\right)
\,.
\ea
%###
%
Consequently, a vanishing superfluid density 
$\partial \langle\hat J\,\rangle /\partial \eta 
=\ord(\eta)$, which marks the end of the
(homogeneous) superfluid phase may occur even when the total 
(thermal plus quantum) depletion is very small 
$|\psi|^2\gg\langle\hat\chi_\beta^\dagger\hat\chi_\beta\rangle$ 
and hence the condensate fraction is still near one.
Note that this behavior is opposite to (bulk) superfluid $^4\!$He,
where the superfluid fraction (near 100\% for low temperatures)
strongly exceeds the condensate part (of order 10\%). 

%%%%%%%%%%%%%%%%%%%%%%%%%%%%%%%%%%%%%%%%%%%%%%%%%%%%%%%%%%%%%%%%%%%%%%%%%%%%%%
%%%%%%%%%%%%%%%%%%%%%%%%%%%%%%%%%%%%%%%%%%%%%%%%%%%%%%%%%%%%%%%%%%%%%%%%%%%%%%
\section{``Supercooled'' states}\label{supercooled}
%%%%%%%%%%%%%%%%%%%%%%%%%%%%%%%%%%%%%%%%%%%%%%%%%%%%%%%%%%%%%%%%%%%%%%%%%%%%%%
%%%%%%%%%%%%%%%%%%%%%%%%%%%%%%%%%%%%%%%%%%%%%%%%%%%%%%%%%%%%%%%%%%%%%%%%%%%%%%

Although the depletion was small 
$|\psi|^2\gg\langle\hat\chi_\beta^\dagger\hat\chi_\beta\rangle$ 
in the cases under consideration, we would like to stress that the
presented controlled 
mean-field expansion (\ref{mean-field}) can also be applied 
to the case of large depletions 
$\langle\hat\chi_\beta^\dagger\hat\chi_\beta\rangle=\ord(|\psi|^2)$. 
This generalization can be achieved by demanding that $V_k$ is
strongly peaked at the origin $V_{k=0}=V_\Sigma = \ord(1)$ 
and much smaller otherwise $V_{|k|>k_0}\ll1$ 
such that the width $k_0$ of the 
peak at the origin is much smaller than the typical $k$-values 
(position $k_*$ and breadth $1/\sqrt{\gamma}$)
associated with the roton-dips
(where $\langle\hat\chi_k^\dagger\hat\chi_k\rangle$ 
yields the major contribution).
To see how this works, let us compare
$\sum_{\beta}V_{\alpha\beta}\hat\chi_\beta$, which must be small 
within our approach, with the depletion 
$\langle\hat\chi_\alpha^\dagger\hat\chi_\alpha\rangle=
\sum_k\langle\hat\chi_k^\dagger\hat\chi_k\rangle/L$.
Calculating the squared norm 
$\langle|\sum_{\beta}V_{\alpha\beta}\hat\chi_\beta|^2\rangle$,  
we get 
$\sum_k|V_k^2|\langle\hat\chi_k^\dagger\hat\chi_k\rangle/L$. 
Similarly, higher orders yield a sum over several wavenumbers 
containing Fourier components at linear combinations of
roton wave-numbers $V_{k\pm k'}$. 
Consequently, all these terms are suppressed even though the quantum
depletion may be large. 
%
%To ensure the smallness of this term
%(while still allowing for a significant depletion), 
%we must assume that $V_k$ is small at the roton minima 
%(where $\langle\hat\chi_k^\dagger\hat\chi_k\rangle$ 
%yields the major contribution).
%
%Furthermore, higher orders yield a sum over several wavenumbers 
%containing Fourier components at linear combinations of
%roton wave-numbers $V_{k\pm k'}$. 
%
%Their suppression could be achieved by demanding that $V_k$ is
%strongly peaked at the origin $V_{k=0}=V_\Sigma = \ord(1)$ 
%and much smaller otherwise $V_{|k|>k_0}\ll \ord(1)$ 
%such that the width $k_0$ of the 
%peak at the origin is much smaller than the typical $k$-values 
%(position $k_*$ and breadth $1/\sqrt{\gamma}$)
%associated with the roton-dips.  
%
%Note the similarity to the concept of weighted operator sums, but here
%in $k$-space; i.e., if many $k$-modes contribute to the quantum
%depletion, it can be approximated by a classical quantity which can be
%large. 
%
%Instead of using the width of the peaks, one could also obtain the
%desired growth of the roton-measure with a large number of
%spatial dimensions and thus many roton dips.

Given these requirements, one may obtain ``supercooled'' states, which
are long-lived superfluid phases in a parameter region where the true
ground state is supersolid. 
In order to demonstrate the main idea, let us consider the following
{\em gedanken} experiment:
We start in the superfluid phase at $T=0$, where 90\% of the particles
are in the condensate $|\psi^2|$ and 10\% in the quantum
depletion $\langle\hat\chi_\alpha^\dagger\hat\chi_\alpha\rangle$.
Now we remove 80\% of the particles (e.g., by a Raman transition
with no momentum transfer incurred) by
decreasing the condensate part $|\psi^2|$ only, i.e., we leave the
modes with $k\neq0$ forming the quantum depletion untouched.
Simultaneously, we increase the interaction strength $V_k$ 
(e.g., via a Feshbach resonance)  such that the product
$|\psi^2|V_k$ remains constant, leaving the quasiparticle spectrum intact. 
After that procedure, half of the remaining particles are in the
condensate $|\psi^2|$ and the other half are in the quantum depletion  
$\langle\hat\chi_\alpha^\dagger\hat\chi_\alpha\rangle=|\psi^2|$. 
These anomalously large quantum fluctuations are caused by the 
increased interaction $V_k$, which is so strong that the true ground
state (with this filling 
$n=|\psi^2|+\langle\hat\chi_\alpha^\dagger\hat\chi_\alpha\rangle$), having
significantly smaller depletion, is supersolid.
However, the immediate transition to the supersolid state is prevented
by the fact that only half the particles are in the condensate.  
Because the quasiparticle modes have the same positive energies as before,
the system is linearly stable. 
Similar to the thermodynamical instability caused by a negative
superfluid density, the decay to the true supersolid ground state is
mediated by the sub-dominant term 
$\sum_{\beta}V_{\alpha\beta}\psi_\beta^*
\langle\hat\chi_\beta\hat\chi_\alpha\rangle$.
Ergo, the predicted ``supercooled'' state is long-lived and thus might
be accessible to an experimental verification. 
 
%%%%%%%%%%%%%%%%%%%%%%%%%%%%%%%%%%%%%%%%%%%%%%%%%%%%%%%%%%%%%%%%%%%%%%%%%%%%%%
%%%%%%%%%%%%%%%%%%%%%%%%%%%%%%%%%%%%%%%%%%%%%%%%%%%%%%%%%%%%%%%%%%%%%%%%%%%%%%
\section{Conclusion}
%%%%%%%%%%%%%%%%%%%%%%%%%%%%%%%%%%%%%%%%%%%%%%%%%%%%%%%%%%%%%%%%%%%%%%%%%%%%%%
%%%%%%%%%%%%%%%%%%%%%%%%%%%%%%%%%%%%%%%%%%%%%%%%%%%%%%%%%%%%%%%%%%%%%%%%%%%%%%

In summary, by means of a controlled expansion into powers of the
small parameter $1/\sqrt{D}$, yielding the mean field $\psi$ plus
(thermal and quantum) fluctuations $\hat\chi_\alpha$, we 
are able to study the impact of these fluctuations onto the
superfluid-supersolid phase transition analytically.  
In addition to the instabilities indicating the end of the
(homogeneous) superfluid phase known from mean field dynamics, 
which occur when the roton dip touches the $\omega=0$ axis, the
fluctuations induce a thermodynamic instability even {\em before}
reaching the classical critical point $\omega_*=0$. 
This breakdown of the homogeneous superfluid is associated with a
negative superfluid density and occurs rather slowly, % \cite{growth}, 
since changes of the mean field $\psi$ induced by fluctuations
$\hat\chi_\alpha$ are governed by the sub-dominant term 
$\sum_{\beta}V_{\alpha\beta}\psi_\beta^*
\langle\hat\chi_\beta\hat\chi_\alpha\rangle$ in Eq.~(\ref{BdG}), which 
is effectively a $\ord(1/\sqrt{D})$-correction to the
Gross-Pitaevski\v\i\/ Eq.~(\ref{GP}). 
%\bibitem{growth} 
%
%The same applies to the thermodynamical instability for $T>0$. ###
%
Finally, even though the thermodynamical instability 
effect is governed by thermal fluctuations, 
quantum fluctuations do also generate intriguing
phenomena near the critical point like supercooled states. 
%\cite{supercooled,Bogoliubov}.

%We find a variety of instabilities indicating the end of the
%(homogeneous) superfluid phase.
%
%For $\eta=0$ and $T=0$, the quasiparticle frequencies $\omega_k$ at
%both roton dips $\pm k_*$ become imaginary beyond the critical point
%and hence these quasiparticle modes start to grow exponentially on
%their own.  
%
%In the presence of a phase gradient $\eta\neq 0$ (but still at $T=0$), 
%on the other hand, the transition occurs earlier and the frequencies
%$\omega_k$ at one roton dip $k_*$ become negative 
%(but not imaginary). 
%
%Hence only the coupling to some environment (fixed by the lattice)
%induces an instability of these quasiparticle modes.
%
%Finally, at finite temperatures $T>0$, the (homogeneous) superfluid
%phase may already become unstable in a region where all the
%quasiparticle modes are still stable, $\omega_k>0$, but the superfluid
%density becomes negative. 
%
%This sequence is ordered in the sense that the first mentioned
%instability is the strongest and occurs last (when coming from the
%superfluid side) etc. 
%
%The last mechanism is rather slow since the change of the
%mean field induced by fluctuations is governed by the sub-dominant
%term $\sum_{\beta}V_{\alpha\beta}\psi_\beta^*
%\langle\hat\chi_\beta\hat\chi_\alpha\rangle$ in Eq.~(\ref{BdG}), which
%is effectively a $\ord(1/\sqrt{D})$-correction to the
%Gross-Pitaevski\v\i\/ Eq.~(\ref{GP}). 

\acknowledgments
%%%%%%%%%%%%%%%%%%%%%%%%%%%%%%%%%%%%%%%%%%%%%%%%%%%%%%%%%%%%%%%%%%%%%%%%%%%%%%

We thank G.\,E. Volovik for helpful discussions. 
This work was supported by the Australian Research Council, the 
Emmy Noether Programme of the German Research Foundation 
(DFG, grant SCHU~1557/1-2,3), as well as the DFG grant FI 690/3-1.

%$^*$\,{\tt schuetz@theory.phy.tu-dresden.de}

%%%%%%%%%%%%%%%%%%%%%%%%%%%%%%%%%%%%%%%%%%%%%%%%%%%%%%%%%%%%%%%%%%%%%%%%%%%%%%


\begin{thebibliography}{9999}
\bibitem{non-adiabatic} 
Note, however, that an opposite current could also be generated by
pure quantum effects due to non-equilibrium phenomena: approaching the
transition too fast, we get non-adiabatic excitations,  
$\langle\hat b_k^\dagger\hat b_k\rangle>0$, which are stronger for
small $\omega_k$.
\end{thebibliography}
\end{document}